\documentstyle[sprocl]{article}

\def\be{\begin{equation}}
\def\ee{\end{equation}}
\def\bea{\begin{eqnarray}}
\def\eea{\end{eqnarray}}

\begin{document}

\title{SEMICLASSICAL THEORY OF {\it h/e} AHARONOV-BOHM OSCILLATION IN BALLISTIC REGIMES}

\author{SHIRO KAWABATA}

\address{Physical Science Division, Electrotechnical Laboratory, Umezono 1-1-4, Tsukuba, Ibaraki 305-8568,
Japan \\E-mail: shiro@etl.go.jp} 




\maketitle\abstracts{ 
We study the magneto-transport in Aharonov-Bohm (AB) billiards forming doubly connected structures.
In these systems, non-averaged conductance oscillates as a function of magnetic flux with period $h/e$.
We derive formulas of the correlation function $C(\Delta \phi)$ of the magneto-conductance for chaotic and regular AB billiards by use of the semiclassical theory.
The different higher harmonics behaviors for $C(\Delta \phi)$ are related to the differing distribution of classical dwelling times.
The AB oscillation in ballistic regimes provides an experimental probe of quantum signatures of classical chaotic and regular dynamics.
}

\section{Introduction}

Electron transport through ballistic quantum billiards is an exceedingly rich experimental system, bearing the quantum signature 
of chaos.\cite{rf:QC}
One of the interesting result that has emerged concerns the magneto-transport of doubly connected ballistic 
billiards, i.e., Aharonov-Bohm (AB) billiards.\cite{rf:Kawabata,rf:Taylor}
We have calculated the $average$
conductance for these systems and showed that the self-averaging effect causes the $h/2e$ 
Altshuler-Aronov-Spivak (AAS) oscillation which is ascribed
to interference between time-reversed coherent back-scattering classical trajectories.\cite{rf:Kawabata} 
Moreover we have showed that the AAS oscillation in these systems
becomes an experimental probe of the quantum chaos. 
Another interesting phenomenon in these systems is the $h/e$ AB oscillation 
for $non-averaged$ conductance.
The result of numerical calculations\cite{rf:Kawabata} indicated that the period of the energy averaged conductance changed from $h/2e$
to $h/e$, when the range of energy average $\Delta E$ is decreased.
However, little is known about the effect of chaos on the $h/e$ AB oscillation in AB billiards.
In this paper, we shall calculate the correlation function $C(\Delta \phi)$ of the $non-averaged$ conductance
by using the semiclassical theory
and show that $C(\Delta \phi)$ is qualitatively different between 
chaotic and regular AB billiards.\cite{rf:Kawabata2}  

\section{Semiclassical Theory}

In the following, we shall derive $C(\Delta \phi)$ separately for chaotic and regular
AB billiards in which uniform normal magnetic 
field $B$ (AB flux) penetrates only through the hollow. 
The  transmission amplitude from a mode $m$ on the left to a mode $n$ on the right
for electrons at the Fermi energy is given by 
\begin{equation}
    t_{n,m}  =  - i \hbar \sqrt{\upsilon_n \upsilon_m} 
    \int dy \int dy' \psi_n^*(y') \psi_m(y) 
  G(y',y,E_F)
  ,
  \label{eqn:e4-4-2}
\end{equation}
where \(\upsilon_m(\upsilon_n)\) and \(\psi_m(\psi_n)\) are the longitudinal velocity and transverse 
wave function for the mode $m$ ($n$) at a pair of lead wires attached to the billiards.  
In eq.~(\ref{eqn:e4-4-2}), $G$ is the retarded Green's function.
In order to carry out the semiclassical approximation, we replace $G$ by the semiclassical Green function, 
\begin{equation}
  G^{sc}(y',y,E) = \frac {2 \pi} {(2 \pi i \hbar)^{3/2}} \sum_{s(y,y')} 
  \sqrt{D_s} 
  \exp \left[ 
                            \frac i {\hbar} S_s (y',y,E) - i \frac \pi {2} \mu_s
       \right]
  \label{eqn:e4-4-4}
\end{equation}
where $S_s$ is the action integral along a classical path $s$,
the pre-exponential factor is
\begin{equation}
  D_s = \frac{ m_e } {\upsilon_F \cos{\theta'}} 
             \left| 
                \left(
			       \frac{\partial  \theta }{\partial y' }
			    \right)_{y}
			 \right|
  \label{eqn:e4-4-5}
\end{equation}
with $\theta$ and $\theta'$ the incoming and outgoing angles, respectively, and $\mu$ is the Maslov 
index.  
Substituting eq.~(\ref{eqn:e4-4-4}) into eq.~(\ref{eqn:e4-4-2})
and carrying out the double integrals by the saddle-point approximation, we obtain  
\begin{equation}
  t_{n,m} = - \frac {\sqrt{2 \pi i \hbar}} {2 W} \sum_{s(\bar n,\bar m)} 
  {\rm sgn} (\bar n) {\rm sgn} (\bar m) \sqrt{\tilde D_s}
   \exp{ 
        \left[
              \frac i {\hbar} \tilde S_s (\bar n,\bar m;E)-i \frac \pi {2} \tilde \mu_s
        \right]
	  }
	  ,
  \label{eqn:e6-2-2}
\end{equation}
where $W$ is the width of the hard-wall leads and \( \bar m = \pm m \).
In eq.~(\ref{eqn:e6-2-2}), 
\( 
  \tilde{S_s} (\bar{n},\bar{m};E) = S_s(y'_0,y_0;E)+ \hbar \pi ( \bar{m} y_0 - \bar{n} y'_0 ) / W 
\), 
\( 
  \tilde{D_s} = ( m_e \upsilon_F \cos{\theta'})^{-1} \left| ( \partial y /\partial \theta' )_{\theta} \right|
\) 
and
\( 
\tilde{\mu_s} = \mu_s + H \left( -( \partial \theta / \partial y )_y' \right)
                      + H \left( -( \partial \theta' / \partial y' )_{\theta} \right),
\)
respectively, where $\theta=\sin^{-1}(\bar{n} \pi / k W )$ and $H$ is the Heaviside step function.

The fluctuations of the conductance $g=(e^2/\pi \hbar) T(k) = (e^2/\pi \hbar) \sum \nolimits _{n,m} \left| t_{n,m} \right|^2 $ are defined by their deviation from the classical
value; in the absence of any symmetries,
%
%
\begin{equation}
 \delta g \equiv g - g_{cl} 
 .
\label{eqn:e6-2-9}
\end{equation}
%
%
In this equation $g_{cl} = (e^2/\pi \hbar) T_{cl}$, where $T_{cl}$ is the classical total transmitted intensity.
In order to characterize the $h/e$ AB oscillation, we define the correlation function of the oscillation in 
magnetic field $B$ by the average over $B$,
%
%
\begin{equation}
 C(\Delta B) \equiv \left< \delta g(B) \delta g(B+\Delta B)\right>_{B} 
 .
\label{eqn:e6-2-10}
\end{equation}
%
%
With use of the ergodic hypothesis, $B$ averaging can be replaced by the $k$ averaging, i.e.,
%
%
\begin{equation}
 C(\Delta B) = \left< \delta g(k,B) \delta g(k,B+\Delta B) \right>_{k} 
 .
\label{eqn:e6-2-11}
\end{equation}
%
%
The semiclassical correlation function of transmission coefficients is given by
%
%
%
%
\begin{eqnarray}
C (\Delta \phi) 
             = 
				  \left( \frac{e^2}{\pi \hbar} \right)^2 &&
                  \frac{1}{8}
				  \left(
				  \frac{\cosh \delta \!-\! 1}{\sinh \delta}
				  \right)^2
				  \cos \left( 2 \pi \frac{\Delta \phi}{\phi_0} \right)        
                  \nonumber\\
			 &\times& 
                  \left\{
                     1 + 2 \sum_{n=1}^{\infty} e^{-\delta n}
			         \cos \left( 2 \pi n \frac{\Delta \phi} {\phi_0} \right)
                  \right\}^2 
				  ,
\label{eqn:e6-2-24}
\end{eqnarray}
%
%
where \( \delta = \sqrt { 2 T_0 \gamma / \alpha } \).\cite{rf:Kawabata2}
In deriving eq.~(\ref{eqn:e6-2-24}) we have used the exponential dwelling time 
distribution, $N(T) \sim \exp(- \gamma T)$,\cite{rf:BJS} and the Gaussian winding number distribution for 
fixed $T$,\cite{rf:Berry} i.e.,
%
%
\begin{equation}
P(w;T)  = \sqrt{ \frac{T_0} {2 \pi \alpha T}} 
     \exp \left( -\frac{w^2 T_0 } { 2 \alpha T} \right)
  \label{eqn:b4}
  ,
\end{equation}
%
%
where $T_0$ and $\alpha$ are the system-dependent constants corresponding to
the dwelling time for the shortest classical winding trajectory and 
the variance of the distribution of $w$, respectively.

On the other hand, for the regular cases, we use $N(T) \sim T^{-\beta}$. \cite{rf:BJS}
Assuming as well the Gaussian distribution of $P(w;T)$, we get
%
%
\begin{eqnarray}
 C (\Delta \phi) &=& 
                  C (0)
 				  \cos \left( 2 \pi \frac{\Delta \phi}{\phi_0} \right)
                  \nonumber\\
                 &\times&
				  \left\{
                     \frac{
                     1+2 \displaystyle{\sum_{n=1}^{\infty}
                      F  \left( \beta-\frac1{2}, \beta+\frac1{2};-\frac{n^2}{2\alpha}  \right)
                      \cos \left( 2 \pi n \frac{\Delta \phi} {\phi_0} \right) }
					  }
					  {
                     1+2  \displaystyle{\sum_{n=1}^{\infty} 
                     F  \left( \beta-\frac1{2}, \beta+\frac1{2};-\frac{n^2}{2\alpha}  \right)}
 					  }                  
				  \right\}^2
				  ,
\label{eqn:e6-2-29}
\end{eqnarray}
%
%
where $F$ is the hyper-geometric function of confluent type. \cite{rf:Kawabata2}

 Next we shall see the difference of $C(\Delta \phi)$ for chaotic and regular AB billiards in detail.
In the chaotic AB billiard, a main contribution to the AB oscillation comes from the $n=1$ component.
On the other hand, for regular cases, the amplitude of the AB oscillation decays algebraically, i.e., $F \sim n^{-2 \beta -1}$
for large $n$. 
This behavior is caused by the power law dwelling time distribution, i.e., $N(T) \sim T^{-\beta}$. 
Thus, in contrast to the chaotic cases, we can expect that the considerably higher harmonics contribution
causes a  noticeable  deviation from the cosine function for $C(\Delta \phi)$.
Therefore, the difference of $C(\Delta \phi)$ of these ballistic AB billiards
can be attributed to the difference between chaotic
and regular classical scattering dynamics.
\section{Summary}
We have investigated that magneto-transport in single ballistic billiards whose structures form AB geometry by use 
of semiclassical methods with a particular emphasis on the derivation of the 
semiclassical formulas. 
The existence of the AB oscillation of $non-averaged$ magneto-conductance is predicted for single chaotic and 
regular AB billiards. 
Furthermore, we find that the difference between  classical dynamics leads to qualitatively different behaviors
for the correlation function. 
The AB oscillation in the ballistic regime will provide a new experimental testing ground for exploring quantum chaos.
\section*{Acknowledgments}

I would like to acknowledge 
K. Nakamura and Y. Takane
for valuable discussions and comments.

\end{document}